**Title:** A mean-variance optimized portfolio constructed for investment in a reference security, for an investor with a preference towards an accepted set of securities.

**Author:** Sidharth Mallik

**Abstract:** We consider a reference security, understood to be an attractive investment, with the caveat that an investor is not willing to directly invest in the security, for presence of constraints, either investor specific or pertaining to the security itself. The investor, however, is open to a portfolio constructed with an accepted set of securities, where returns could be considered similar to the reference security. We demonstrate, under a measure of similarity, such a portfolio could be selected with a mean-variance characterization, as defined by Markowitz. Furthermore, we consider the performance relative to the reference security, with the Sharpe Ratio. The objective of the paper is to derive an optimal portfolio to address an investor preference for the accepted set of securities.

**Key Messages:**
- Investment preference for an accepted set of securities.
- A mean-variance optimized portfolio for the considered measure of similarity.
- A set of tests to decide how similar the optimal portfolio is to the reference security.

**Main text:**

**Introduction:** We introduce the portfolio selection problem as described by Markowitz (1952), with a detailed explanation in Markowitz (1991). We highlight notable aspects as statements,
<u>Statement 1</u>, "a portfolio analysis starts with information concerning individual securities, and the purpose is to find what best meets the objectives of the investor."
<u>Statement 2</u>, "the choice of analysis depends on the nature and goals of the investor."
<u>Statement 3</u>, "alternative inputs could be considered as part of the analysis."
<u>Statement 4</u>, "a portfolio is considered efficient if it is not possible to obtain higher expected return without admitting greater variability."
     In this paper, we consider a portfolio analysis problem whereby an investor finds a security attractive but is unwilling to directly invest. Mean and variance are considered measures for expected return and variability, respectively. The corresponding optimization problem is then solved to obtain efficient portfolios, termed as optimal portfolios. Furthermore, we consider a requirement for assessing the similarity of such a portfolio relative to the reference security. We design two tests, a Student's t-test and a test derived from the Sharpe Ratio, as defined by Sharpe (1994). We demonstrate, for the given investor preferences, the Sharpe Ratio is derived to have a theoretical value, and hence, could be tested for a decision on similarity.

**Investment with a reference security:** We consider an investor who finds a security attractive. However, because of constraints, either on the part of the investor, or pertaining to the security itself, the investor decides not to invest directly. Such a security is referred to as a reference security. While this may appear to be non-obvious at first, examples do exist, as has been illustrated in a following section.

**Accepted set of securities:** The investor, however, is open to consider a portfolio with a performance similar to the reference security. We interpret this as investing in the reference security through a portfolio of an accepted set of securities, not including the reference security. The accepted set of securities may or may not be related with the reference security or the concerned sector. Furthermore, the accepted set of securities are allowed to have a positive, negative or a zero correlation with the reference security.

We illustrate two examples where such an investor preference could exist.

**Example 1:** A recent example is one of carbon markets. Instruments related to carbon markets have been listed in exchanges. Studies like Zhang *et al* (2017) have suggested benefits of considering such an investment. However, for reasons of novelty or for lack of interest or simply a wait and watch approach, a section of investors has not participated in the instruments. The investors, however, have invested in companies having a direct investment in instruments related to carbon markets, implying an investment on part of the investors. This may be thought of as a situation where an instrument in carbon markets could be considered as a reference security with the equities as the accepted set.

**Example 2:** For this example, we consider an investor who has decided to hold onto an existing set of securities, whereby returns have been favourable in the past. However, due to an innovation, another investment has become attractive. The innovation impacts the existing set of securities in the investor's portfolio, the companies in the portfolio having an investment themselves in the innovation. The investor, however, for the decision to hold onto the existing set, does not want to include fresh listings in the portfolio. Such a listing could then be considered as a reference security, while the set of favourable securities is the accepted set for the investor.

**Expected returns of the reference security:** We further consider the investor to prefer expected returns of the reference security. There is a rich body of literature revolving around investment with expected returns. We refer two studies in this regard, of Jegadeesh (1990) and Fama and French (1992). To understand this, the investor, having observed the returns of the reference security, decides on pursuing expected return (calculated to be the average) as a benchmark for the portfolio's returns. A point in favour of such a decision is, an expected return could be estimated at a required instant of time with a confidence interval. We do not consider the portfolio to be mimicking the reference security exactly, but only to be similar in the sense defined. A point to note here is with regards to the sample estimate of expected return of the reference security, where the confidence interval depends on the variance of the security. Another point to note is with regards to availability of data to arrive at the estimate of expected return. This may entail an additional cost to the investor and hence could be adjusted in the estimate. We refer to this as a cost of monitoring the reference security.

**Notation:** For this paper, we do not direct our attention to a particular set of securities. Hence, we make a generalised assumption of normally distributed returns. We denote the one-period returns of the reference security as $r_{rs}$, distributed normally with mean $\mu_{rs}$ and variance $\sigma_{rs}^2$. We denote $N$ as the number of securities in the accepted set. Returns of the *ith* security is denoted as $r_i$, distributed normally with mean $\mu_i$ and variance $\sigma_{ii}^2$. The corresponding covariances are denoted as $\sigma_{ij}^2$ for the *ith* and the *jth* security. A portfolio is a combination of securities from the accepted set. The weight of the *ith* security in a portfolio is denoted as $w_i$. The returns of the portfolio are denoted as $r_{po}$. We denote $E[.]$ as expectation and $Var[.]$ as variance. Time intervals at which the portfolio is active, are indexed by $t$, where $t$ ranges from $1$ to $T$.

**Criterion of similarity:** We consider the portfolio return at time interval $t$ as $r_{po}^{(t)}$. We define the quantity $S$, as the average of sum of squares of the difference of portfolio returns from expected return of the reference security,

$$S = \frac{\sum_{t=1}^{T} (r_{po}^{(t)} - \mu_{rs})^2}{T}$$

The measure $S$ is the average squared deviation of the portfolio from expected return of the reference security. Lower the value of $S$, greater is the similarity. Given a choice, the investor would hence admit a portfolio with a lower $S$. With this measure, the optimal portfolio could be described as the one minimizing $S$. Hence the objective of portfolio optimization could thus be defined as,

$$\text{Minimize } S = \frac{\sum_{t=1}^{T} (r_{po}^{(t)} - \mu_{rs})^2}{T}$$

A portfolio satisfying such a criterion is referred to as an optimal portfolio.

**Deriving the model:** We now consider the returns of the optimal portfolio as $r_{po}^*$. With $r_{po}^{*(t)}$ as independent and identically distributed, the definition of $S$ is the variance of returns $r_{po}^*$ with mean $\mu_{rs}$. Hence, if an optimal portfolio does exist, the value of $S$ for such a portfolio could be expressed as,

$$S^* = \frac{\sum_{t=1}^{T} (r_{po}^*(t) - \mu_{rs})^2}{T}$$

where $S^*$ is the value corresponding to the optimal portfolio. $r_{po}^*$ could thus be expressed as the solution to,

$$\text{Minimize } Var[r_{po}]$$
$$\text{subject to } E[r_{po}] = \mu_{rs}$$

**The constraint on weights:** We allow another constraint to be added. As the investor considers the optimal portfolio and the reference security to be similar, a unit of the optimal portfolio matches a unit of the reference security. In terms of weights of securities within the portfolio, this is represented as,

$$\sum_{i=1}^{N} w_i = 1$$

Furthermore, we consider the investor is open to short sales, thus allowing negative values for $w_i$. Additional constraints could be added, specific to an investor as considered by Wanka and Göhler (2001), Jacobs *et al* (2004) and Jacobs and Levy (2012).

**Procedure of optimization:** We summarize the model of optimization hence developed as,

$$\text{Minimize } Var[r_{po}]$$
$$\text{subject to } E[r_{po}] = \mu_{rs}$$
$$\text{and } \sum_{i=1}^{N} w_i = 1$$

To solve the optimization problem, we apply the Method of Lagrange Multipliers. We express the variance of a portfolio as,

$$\sum_{i=1}^{N} \sum_{j=1}^{N} w_i w_j \sigma_{ij}^2$$

and expected return as,

$$\sum_{i=1}^{N} w_i \mu_i$$

We, thus represent the optimization problem as,

$$\text{Minimize } \sum_{i=1}^{N} \sum_{j=1}^{N} w_i w_j \sigma_{ij}^2$$
$$\text{subject to } \sum_{i=1}^{N} w_i \mu_i = \mu_{rs}$$
$$\text{and } \sum_{i=1}^{N} w_i = 1$$

The Lagrangian is,

$$L = \sum_{i=1}^{N} \sum_{j=1}^{N} w_i w_j \sigma_{ij}^2 - \lambda_1 (\sum_{i=1}^{N} w_i \mu_i - \mu_{rs}) - \lambda_2 (\sum_{i=1}^{N} w_i - 1)$$

We then equate the following derivatives to 0,

$$\frac{\partial L}{\partial w_i} = 2(\sum_{j=1}^{N} w_j \sigma_{ij}^2) - \lambda_1 (\mu_i) - \lambda_2 = 0$$

$$\frac{\partial L}{\partial \lambda_1} = -(\sum_{i=1}^{N} w_i \mu_i - \mu_{rs}) = 0$$

$$\frac{\partial L}{\partial \lambda_2} = -(\sum_{i=1}^{N} w_i - 1) = 0$$

We arrive at a system of linear equations,

$$AX = b$$

where,
$A$: The matrix of coefficients.
$X$: The vector consisting of weights $w_i$ along with $\lambda_1$ and $\lambda_2$.
$b$: The vector of constants in the equations, consisting of 0's along with $\mu_{rs}$ and 1.
For a non-singular matrix of coefficients, we could solve the equation as,

$$X = A^{-1} b$$

The solution thus obtained, provides us the values of $w_i$ as an extremum of the objective function. On constructing the Hessian matrix $H$ of the objective function, we obtain,

$$H = [h_{ij}] = [2(\sigma_{ij}^2)]$$

This being a multiple of the variance-covariance matrix of securities, is positive definite. Thus, the extremum found corresponds to a minimum of the objective function. The procedure, thus finds the weights corresponding to the optimal portfolio.

**Measuring similarity:** Once the optimal portfolio has been found, tests to decide the similarity of the portfolio could be desirable for the investor. We consider two such tests, both leading to respective hypotheses. The first is a Student's t-test for expected return of the portfolio, while the second is derived from the Sharpe Ratio.

**Student's t-test for the hypothesis of expected returns:** Given that we considered returns to be normally distributed, and the assumption of expected return of the optimal portfolio to be equal to expected return of the reference security, we perform the Student's t-test to check for the hypothesis of mean of returns of the optimal portfolio to be equal to mean of returns of the reference security. The null hypothesis is thus defined as,

$H_0$: The mean of returns of the optimal portfolio equals the mean of returns of the reference security

Confidence intervals are constructed with the Student's t-distribution and $H_0$ is accepted or rejected accordingly.

**Sharpe Ratio for the optimal portfolio with the reference security:** We derive an estimate along the lines of the definition of Sharpe Ratio for the optimal portfolio with the reference security. For this, we construct a series of differences between the returns, defined as,

$$r_{diff} = r_{rs} - r_{po}^*$$

The expected value of $r_{diff}$ could be expressed as,

$$E[r_{diff}] = E[r_{rs} - r_{po}^*]$$
$$\text{or, } E[r_{diff}] = E[r_{rs}] - E[r_{po}^*]$$
$$\text{implies, } E[r_{diff}] = \mu_{rs} - \mu_{rs} = 0$$

The standard deviation of $r_{diff}$ is positive unless $r_{diff}$ is identically 0 at all time intervals, a possibility we consider as exceptional. By definition, therefore, the Sharpe Ratio is estimated as,

$$\text{Sharpe Ratio} = \frac{\text{mean}(r_{diff})}{\text{standard deviation}(r_{diff})}$$
$$\text{or, Sharpe Ratio} = 0$$

We therefore arrive at a definite value for the Sharpe Ratio, equal to 0. A test of hypothesis with the Student's t-distribution, the null hypothesis being 0, gives a method to decide how similar the optimal portfolio has been relative to the reference security. We express the test as,

$H_0$: Sharpe Ratio for $r_{diff}$ equals to 0

Once again, confidence intervals are constructed with the Student's t-distribution and $H_0$ is accepted or rejected accordingly. This is different from applying the Sharpe Ratio to estimate the outperformance of a security relative to another. For the optimal portfolio, closer the value to 0, more similar is the performance relative to the reference security.

**Conclusion:** From the model hence derived, we arrive at the following conclusions,
1. Under the mean-variance characterization, defined by Markowitz, the optimal portfolio is considered similar to the reference security.
2. The optimal portfolio may also be seen as an efficient portfolio, whereby the variance is minimum for a given expected return.

3. The derived value of the Sharpe Ratio as equal to 0, lets us test for the similarity. For accepted values of the Sharpe Ratio, the optimal portfolio could be considered as an investment for the defined investor preferences.

**Acknowledgements:**
1. We acknowledge availing the facilities provided by University of Hull. This includes access to journals for reference.

**Declarations of Interest:** The author reports no conflict of interest. The author alone is responsible for the content and writing of the paper. Aspects of writing style could be seen as updated from previous paper(s) of the author, where applicable.